\documentclass{article}

\usepackage{amsmath}
\usepackage{tikz} 
\usetikzlibrary{arrows}
\usepackage{url}
\usepackage{amsfonts,amssymb, booktabs} 
\usepackage{mathtools}
\usepackage{longtable} 
\usepackage{float}
\usepackage{placeins}
\usepackage{graphicx}
\usepackage{comment}

\usepackage{afterpage}
\usepackage{emptypage}

\usepackage{algorithm}
\usepackage{algorithmic}
\usepackage{subcaption}
\usepackage{chngcntr}
\usepackage{comment}

\usepackage[utf8]{inputenc}

\def\indep{\perp\hskip -9pt \perp }

\title{Covariate selection for the estimation of marginal hazard ratios in high-dimensional data}

\usepackage{authblk}

\author[]{Guilherme W. F. Barros\thanks{Corresponding author: guilherme.barros@umu.se}}
\author[]{Jenny H\"aggstr\"om}
\affil[]{Department of Statistics, Ume{\aa} School of Business, Economics and Statistics, Ume{\aa} University, SE-901 87 Ume{\aa}, Sweden}

\usepackage{csquotes}
\usepackage{booktabs}
\usepackage{multirow}
\usepackage[margin=2cm]{geometry}
\usepackage{longtable}
\usepackage{comment}
\usepackage{hyperref}
\usepackage[
backend=biber,
style=authortitle,
citestyle=authoryear,
maxbibnames=100,
natbib,
defernumbers=true,
dashed = FALSE
]{biblatex}
\renewbibmacro{in:}{}
\DeclareFieldFormat[article, inbook, incollection, inproceedings, misc, thesis, unpublished]
{title}{#1}
\DeclareFieldFormat[article]
{pages}{#1}
\DeclareBibliographyCategory{cited}
\AtEveryCitekey{\addtocategory{cited}{\thefield{entrykey}}}

\newcommand{\commenttext}[1]{}

\begin{document}
\renewcommand{\arraystretch}{1.2}
\maketitle

\begin{abstract}
\noindent
Hazard ratios are frequently reported in time-to-event and epidemiological studies to assess treatment effects. In observational studies, the combination of propensity score weights with the Cox proportional hazards model facilitates the estimation of the marginal hazard ratio (MHR). The methods for estimating MHR are analogous to those employed for estimating common causal parameters, such as the average treatment effect. However, MHR estimation in the context of high-dimensional data remain unexplored. This paper seeks to address this gap through a simulation study that consider variable selection methods from causal inference combined with a recently proposed multiply robust approach for MHR estimation. Additionally, a case study utilizing stroke register data is conducted to demonstrate the application of these methods. The results from the simulation study indicate that the double selection covariate selection method is preferable to several other strategies when estimating MHR. Nevertheless, the estimation can be further improved by employing the multiply robust approach to the set of propensity score models obtained during the double selection process.

\end{abstract}


\section{Introduction}
\label{section:Introduction}

In time-to-event studies, hazard ratios are commonly estimated to investigate the effects of interventions and treatments. Here, the focus is on the marginal hazard ratio (MHR) which quantifies the population-averaged over time effect of a binary treatment on a time-to-event outcome.

In randomized controlled trials (RCTs) MHR can be estimated by fitting a Cox proportional hazards model that includes only the treatment variable. However, such a direct approach is not appropriate in observational studies due to the potential for confounder-induced bias \citep{austin_best_practice}. 

It has been shown that MHR can be estimated in observational studies by initially balancing the covariates of the treated and the untreated groups \citep{austin-3, austin_fullmatching_misspecification}. This can be achieved using propensity score methods, aiming to reduce systematic differences between the groups, thereby mimicking the conditions of an RCT  \citep{austin_diff_methods}. Once the sample is balanced, MHR can be estimated using the same procedure as in an RCT. These weighting methods used to estimate MHR are  analogous to causal inference methods used to estimate parameters such as the average treatment effect (ATE) and the average treatment on the treated (ATT) \citep{book_whatif}.

For unbiased estimation of MHR, the propensity score model and the corresponding weights must be correctly specified \citep{austin_fullmatching_misspecification}. To address this requirement, there has been recent developments in methods for weight estimation using a multiply robust approach. This approach has proven effective, provided that at least one model within the postulated set is correctly specified \citep{multi_robust}.

Furthermore, observational studies often face the challenge of high-dimensional covariate spaces, such as those encountered in large-scale register studies or genetic data research, and the high-dimensionality is often exacerbated due to the inclusion of  higher order moments and interaction terms in the propensity score model. Consequently, variable selection has been of interest in the causal inference literature, where the primary goal is identifying a set that fully accounts for all confounding \citep{causal_inf_gao, causal_inf_schneeweiss, causal_inf_vanderWeele, causal_inf_hill, selection_de_luna, selection_haggstrom, selection_persson}. There are some important factors to consider about such a set: variables that are causes of the outcome are known to increase precision, while variables that are solely causes of the treatment (instruments) decrease it \citep{causal_inf_brookhart}. In a high-dimensional non-time-to-event outcome setting, the post-double-selection estimator has been shown to result in uniformly valid inference \citep{ds_belloni}, but at the cost of larger standard errors compared to estimators selecting away instruments \citep{selection_moosavi}.

To the best of the authors' knowledge, there have been no studies on best practices for MHR estimation in observational, high-dimensional contexts. This paper aims to bridge this gap by evaluating various approaches to variable selection pertinent to the aforementioned contexts. The paper is organized as follows: Section \ref{section:Methodology} reviews MHR estimation using propensity score weights, covariate selection strategies and, multiply robust weight estimation. In Section \ref{section:Simulation} the simulation design and results are described. The same methods studied in simulations are then applied to a case study, utilizing stroke register data, in Section \ref{section:case_study}. In Section \ref{section:discussion}, the paper is concluded with a short discussion of the results, suggestions for future work and recommendations for applied researchers . 

\section{Methods}
\label{section:Methodology}

Suppose for each individual $i = 1, \ldots, n$ we have $P$ measured covariates $\mathbf{X}_{i} = (X_{i1}, \ldots, X_{iP})^T$. We define $Z_i$ as the treatment status ($Z_i = 1$ if treated; $Z_i = 0$ if untreated) and $T_i = \min\{Y_i, C_i\}$ is the observed outcome, where $Y_i$ is the true time-to-event and $C_i$ is the censoring time. $D_i = 1_{Y_i\leq C_i}$ is the event indicator. We assume the censoring to be non-informative, that is, the distribution of the time-to-event of the individuals provides no information on their censoring times, and vice versa. Furthermore, we assume unconfoundedness, i.e., $\{Y_i(0), Y_i(1)\} \indep T|\mathbf{X}_i$.

Consider the following Cox proportional hazards model 

\begin{equation}
\label{equation:cph}
h(t|Z) = h_0 (t)\mbox{exp}(\alpha_Z Z),
\end{equation}
where $h_0(t)$ is the hazard function when $Z = 0$ and $\mathbf{X}$ is marginalized and $\mbox{exp}(\alpha_Z)$ is MHR, i.e., the hazard ratio between $Z = 1$ and $Z = 0$ when $\mathbf{X}$ is marginalized. In a perfect RCT without censoring, fitting \ref{equation:cph} to the  data would result in an unbiased estimate of MHR \citep{wyss}. However, in the case of observational data, fitting \ref{equation:cph} results in bias due to confounding. In order to address this source of bias, it is possible to use inverse probability of treatment weighting (IPTW) \citep[][Chapter 2]{book_whatif} together with Cox regression to estimate MHR with negligible bias \citep{austin_diff_methods, fireman}. The weights are defined as

\begin{equation}
\label{equation:w_iptw}
w_i^{PS} = \frac{Z_i}{\mbox{Pr}(Z_i = 1|\mathbf{X}_i)} + \frac{1 - Z_i}{1 - \mbox{Pr}(Z_i = 1|\mathbf{X}_i)},
\end{equation}
\noindent
where $\mbox{Pr}(Z_i = 1|\mathbf{X}_i)$ is the propensity score of an individual $i$.

Propensity score weighting was originally proposed for estimating causal parameters in a non-time-to-event context and the weights in \ref{equation:w_iptw} (commonly referred to as ATE weights) are used to estimate the ATE in the total population of both treated and untreated. Typically, the propensity score is estimated using a logistic regression

\begin{equation*}
\mbox{Pr}(Z_i = 1|\mathbf{X}_i) = \frac{\mbox{exp}(\beta_0 + \beta_1 X_{i1} + \dots \beta_P X_{iP})}{1 + \mbox{exp}(\beta_0 + \beta_1 X_{i1} + \dots \beta_P X_{iP})}.
\end{equation*}
\noindent

In the presence of high-dimensional data, logistic regression can be adapted with $\ell_1$ regularization, lasso regression \citep{book_elements}, which puts the following constraint on the parameters

\begin{equation*}
\sum_{p = 1}^P {|\beta_p| \le \lambda_Z},
\end{equation*}
\noindent
\noindent
where $\lambda_Z > 0$ is a tuning parameter that controls the regularization. This regularization induces shrinkage of the estimated parameters towards zero, potentially setting some coefficients exactly to zero, thereby imposing sparsity on  $\boldsymbol{\beta}$. Hence, this approach can be used for variable selection. Higher values of  $\lambda_Z$ result in stronger regularization, which reduces model complexity. Analogous to its application in logistic regression, lasso regression has been adapted for the time-to-event setting \citep{tibshirani_survival_lasso}, where the parameters in the Cox proportional hazards model

\begin{equation*}
h(t|\mathbf{X}_i) = \lambda_0 (t)\mbox{exp}\big(\sum_{p=1}^P \mathbf{X}_{ip} \alpha_p\big).
\end{equation*}

\noindent
are constrained to

\begin{equation*}
\sum_{p = 1}^P {|\alpha_p| \le \lambda_Y}.
\end{equation*}
\noindent

Due to its variable selection properties, lasso regression has been widely used in high-dimensional time-to-event settings to estimate conditional hazard ratios \citep{survival_high_dim_Lasso_application_xu, survival_high_dim_guide_salerno, survival_high_dim_waldron, survival_high_dim_witten}. It is also a well-established method for variable selection in estimating treatment effects within the field of causal inference \citep{lasso_causal_inf_ju, lasso_causal_inf_shortreed, causal_inf_antonelli, causal_inf_gao, causal_inf_avagyan}.

When selecting covariates to control for, we initially consider two sets: those that predict treatment assignment, $X_Z$, and  those that predict the outcome, $X_Y$ \citep[see][for formal definitions]{selection_de_luna}, and their lasso selected counterparts, $\hat{X}_Z$ and $\hat{X}_Y$. In addition, we consider the double selection approach proposed by \citet{ds_belloni}, resulting in the set $\hat{X}_{DS} = \hat{X}_Z \cup \hat{X}_Y$.

Within the causal inference framework, double selection in conjunction with doubly robust estimators of causal parameters has been shown to yield uniformly valid confidence intervals \citep{ds_belloni, selection_moosavi}. Moreover, this post-double-selection estimator has been demonstrated to outperform other strategies in terms of bias and root mean-squared error (RMSE) when estimating ATE in high-dimensional settings \citep{causal_inf_gao}.

For estimation of MHR, another recent advancement is a multiply robust estimation approach adapted to the time-to-event case by \citet{multi_robust}. This method, originally proposed by \citet{multi_robust_original} for estimation of causal parameters such as ATE, provides a consistent estimator of MHR as long as one of the postulated models is correct \citep{multi_robust}. Let $\mathcal{E} = \{e^j(\boldsymbol{\beta}^{j}|\mathbf{X})\}$ be a set of $j = 1, \dots, J$ postulated propensity score models, with $\hat{\boldsymbol{\beta}}^j$ representing the estimated parameters for the $j$th model. We then define the mean of each fitted propensity score,

\begin{equation*}
\hat{\mu}^j = n^{-1}\sum^{n}_{i=1} e^{j}(\hat{\boldsymbol{\beta}}^{j}|\mathbf{X}_{i}),
\end{equation*}
\noindent
and each individual's distance from the $j$th mean, $j = 1, \ldots, J$,

\begin{equation*}
\hat{g_i}(\hat{\boldsymbol{\beta}}) = \left( e^1(\hat{\boldsymbol{\beta}}^{1}|\mathbf{X}_i) - \hat{\mu}^1,\dots, e^J(\hat{\boldsymbol{\beta}}^{J}|\mathbf{X}_i) - \hat{\mu}^J \right).
\end{equation*}
\noindent
Then the weights for the treated individuals $i = 1, \dots, m$ are given by

\begin{equation*}
\hat{\omega}_i = \left( \frac{1}{1 + \hat{\boldsymbol{\rho}}^T \hat{g}_i(\hat{\boldsymbol{\beta}})}\right) \bigg/ m,
\end{equation*}
\noindent
where $\hat{\boldsymbol{\rho}}$ is a vector of size $J$ obtained by solving

\begin{equation*}
\sum^{m}_{i=1} \frac{\hat{g}_i(\hat{\boldsymbol{\beta}})}{1 + \boldsymbol{\rho}^T \hat{g}_i(\hat{\boldsymbol{\beta}})} = \mathbf{0}, 
\end{equation*}
\noindent
using convex minimization which guarantees a unique solution \citep{convex_han}. Analogously, the weights for the untreated individuals $i = m+1, \dots, n$ are given by

\begin{equation*}
\hat{\omega}_i = \left( \frac{1}{1 - \hat{\boldsymbol{\nu}}^T \hat{g}_i(\hat{\beta})}\right) \bigg/ (n-m),
\end{equation*}
\noindent
where $\hat{\boldsymbol{\nu}}$ is a vector of size $J$ obtained by solving

\begin{equation*}
\sum^{n}_{i=m+1} \frac{\hat{g}_i(\hat{\boldsymbol{\beta}})}{1 - \boldsymbol{\nu}^T \hat{g}_i(\hat{\boldsymbol{\beta}})} = \mathbf{0}.
\end{equation*}
\noindent
In our high-dimensional setting, the combinatorial nature of selecting different sets leads to an excessively large number of potential propensity score models, and using the multiply robust approach with such an amount of models is not feasible. To address this issue, we propose using four models generated by the sets of covariates $\hat{X}_Y$, $\hat{X}_Z$, $\hat{X}_{DS}$ and $\hat{X}_I = \hat{X}_Y \cap \hat{X}_Z$, instead of an exhaustive enumeration of all possible sets. The idea behind this is to limit the multiply robust approach to the two simplest sets that can be obtained from regressions on the data and the operations between the two sets. The set of variables predictive of treatment, $\hat{X}_Z$, frequently used in medical research, is increasingly recognized as unsuitable due to the risk of instrument inclusion, leading to variance inflation \citep[see][and references therein]{selection_moosavi}. In non-time-to-event contexts, the set of variables predictive of outcome, $\hat{X}_Y$, has been shown to result in lower variance compared to other sets but may not yield valid inference \citep{selection_moosavi}. By incorporating the union, $\hat{X}_{DS}$, we aim to mitigate any bias arising from omitting a specific confounder in $\hat{X}_Z$ or $\hat{X}_Y$, which is a risk if a confounder is weakly related with either the outcome or treatment. Including the intersection, $\hat{X}_I$, targets the set of true confounders.

\section{Monte Carlo Simulation}
\label{section:Simulation}
A simulation study was conducted to evaluate the usage of lasso for selecting the target covariate sets $X_Z$, $X_Y$, $X_Z\cup X_Y$ and $X_Z\cap X_Y$ within a time-to-event context and to examine how the selected sets impacts the estimation of MHR, both when used separately and as inputs for the multiply robust approach. 

\subsection{Simulation Design}
The simulation design was based on previous simulation studies involving time-to-event data \citep{austin_diff_methods, austin_fullmatching_misspecification, misspec_orig, barros_1}. We generated a $P$-dimensional covariate vector $\mathbf{X}$ where $X_{2+q}$, $X_{4+q}$, $X_{7+q}$ and $X_{10+q}$, $q = 0, 10, 20, \ldots, P-10$, are standard normally distributed and the remaining covariates $X_{1+q}$, $X_{3+q}$, $X_{5+q}$, $X_{6+q}$, $X_{8+q}$, and $X_{9+q}$ follow a Bernoulli(0.5) distribution.  Each individual's true propensity score was generated using a logit function, $\text{logit}(\mbox{Pr}(Z = 1|\mathbf{X}_i)) = \boldsymbol{\beta}^T\mathbf{X}_i$, where $\boldsymbol{\beta} = k \times (0.8, -0.25, 0.6, -0.4, -0.8, -0.5, 0.7, 0, \ldots, 0)^T$. Here, $k$ controls the degree of overlap between the treated and untreated groups; $k = 1$ represents strong overlap, which decreases as $k$ increases. For insights on how $k$ affects overlap, see Figure \ref{fig:k_example} in the Appendix.

The time-to-event of each individual was generated as $Y_i = \left( \frac{-\text{log}(u_i)}{\gamma \text{exp}(LP_i)} \right)^{1/\eta}$, where $u_i$ was sampled from a standard uniform distribution, and the linear predictor (LP) is $LP_i = \alpha_{z}^* Z_i + \boldsymbol{\alpha}^T\mathbf{X}_i$. The parameters $\eta$, $\gamma$, and $\boldsymbol{\alpha}$ were set to 2, 0.00002 and $(0.3, -0.36, -0.73, -0.2, 0, 0, 0, 0.71, -0.19, 0.26, 0, \ldots, 0)^T$, respectively. Hence, only the first ten covariates are related to treatment and/or outcome and the rest are irrelevant.  To achieve a specific MHR, a corresponding value for $\alpha_{z}^*$ must be chosen and this was done with a bissection method. See \cite{austin_diff_methods} and \cite{austin_fullmatching_misspecification} for details about the bissection method and the simulation scenario. 

To account for censoring, we generated censoring times according to  $C_i \sim \mbox{Uniform}(0, \theta)$, resulting in non-informative censoring, a common assumption in time-to-event studies \citep{censoring_types}. The desired censoring rate was achieved by selecting $\theta$ as suggested by \citet{Wan2016}.

The propensity score was estimated using logistic regression, and weights were generated as in \ref{equation:w_iptw}. The weights were then included in a Cox regression as in \ref{equation:cph} to estimate MHR. For estimating the propensity score, we considered the following variable sets: the set of variables affecting treatment assignment $X_Z = \{X_1, \dots, X_7\}$, the set of variables affecting the outcome $X_Y = \{X_1, X_2, X_3, X_4, X_8, X_9, X_{10}\}$, the union set $X_{Z} \cup X_Y = \{X_1, \dots, X_{10}\}$, lasso selected versions of the aforementioned sets, true confounders $X_{Z} \cap X_Y = \{X_1, X_2, X_3, X_4 \}$ and the set of all variables $X_{all} = \{X_1, \ldots, X_P\}$. The set $\hat{X}_Z$ comprises covariates associated with non-zero coefficients resulting from a lasso regression using all covariates with the treatment variable $Z$ as response. To obtain $\hat{X}_Y$, two survival lasso regression are executed: one for the treated group, resulting in the set $\hat{X}_{Y(1)}$, and another for the untreated group, resulting in  $\hat{X}_{Y(0)}$. The set $\hat{X}_Y$ is the union of $\hat{X}_{Y(1)}$ and $\hat{X}_{Y(0)}$. For all lasso regressions, the hyperparameters $\lambda_Z$ and $\lambda_Y$ were selected according to 10-fold cross-validation with deviance loss for the logistic regression and Harrel's concordance for the survival case. The value of a hyperparameter chosen after this process is the largest such that the loss is within one standard error of the minimum. Finally, $\hat{X}_{DS} = \hat{X}_Z \cup \hat{X}_Y$ and $\hat{X}_{Rob}$ is the result of the multiply robust approach including the sets $\hat{X}_Z$, $\hat{X}_Y$, $\hat{X}_{DS}$, and $\hat{X}_I = \hat{X}_Z \cap \hat{X}_Y$.

We considered various settings including $n<P$, $n=P$, $n>P$, with censoring, weak overlap ($k = 3$), and cases where MHR is either above or below 1.  A summary of these settings is presented in Table \ref{table:settings}. 

\begin{table}[ht]
\centering
\caption{Summary of the simulation settings}
\begin{tabular}{lrrrrr}
  \hline
Scenario & $n$ & $P$ & $k$ & MHR & Censoring Rate \\ 
  \hline
1 & 1000 & 500 & 1 & 2 & 0\\ 
  2 & 1000 & 1000 & 1 & 2 & 0\\ 
  3 & 1000 & 1500 & 1 & 2 & 0\\ 
  4 & 1000 & 1000 & 1 & 0.5 & 0\\ 
  5 & 1000 & 1000 & 1 & 2 & 0.2\\ 
  6 & 1000 & 1000 & 3 & 2 & 0\\ 
7 & 500 & 250 & 1 & 2 & 0\\ 
  8 & 500 & 500 & 1 & 2 & 0\\ 
  9 & 500 & 750 & 1 & 2 & 0\\ 
  10 & 500 & 500 & 1 & 0.5 & 0\\ 
  11 & 500 & 500 & 1 & 2 & 0.2\\ 
  12 & 500 & 500 & 3 & 2 & 0\\ 
   \hline
\end{tabular}
\label{table:settings}
\end{table}

\noindent
We used 1000 replicates in each scenario. For each of variable selection approach, we computed the mean of the estimated MHR as $\overline{\mbox{MHR}}=\frac{1}{1000}\sum_{r = 1}^{1000} \mbox{exp}(\hat{\alpha}_{Zr})$, the bias as $\mbox{Bias} = \overline{\mbox{MHR}} - \mbox{exp}(\alpha_Z)$, the relative bias as $\mbox{Rel. Bias} = \mbox{Bias}/\mbox{exp}(\alpha_Z)$, the Monte Carlo standard error as $\mbox{SD} =\frac{1}{1000 - 1}\sum_{r = 1}^{1000} (\mbox{exp}(\hat{\alpha}_{Zr}) - \overline{\mbox{MHR}})^2$, and the RMSE as $\mbox{RMSE} = \sqrt{\mbox{Bias}^2 + \mbox{SD}^2}$. A robust sandwich type variance estimator 
 \citep[adequate when considering MHR estimation but known to result in slightly conservative confidence intervals][]{austin-variance} was used to generate 95\% confidence intervals for each MHR estimate and coverage was calculated as the proportion of intervals that contained the true MHR. 

Data generation and all estimations were carried out using the software \texttt{R} \citep{R}. The \texttt{survival} package \citep{survival} was used for Cox modeling. Both the lasso and the cross-validation procedures were implemented via the \texttt{cv.glmnet} function from the \texttt{glmnet} package \citep{glmnet}.

\subsection{Simulation Results}
\label{section:results}

\begin{figure}[htp]
\centering

\begin{subfigure}{0.45\columnwidth}
\centering
\includegraphics[width=\textwidth]{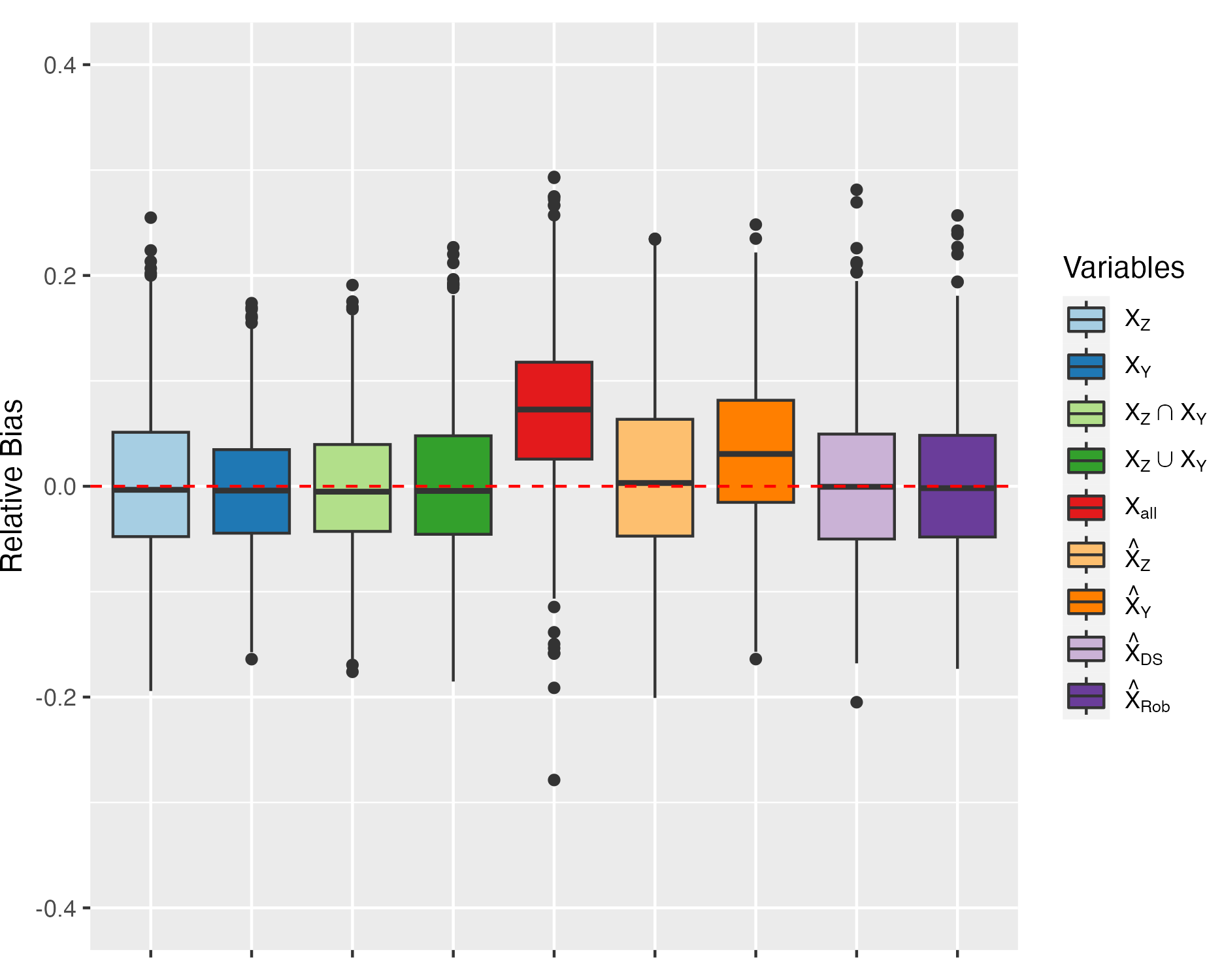}
\caption{Relative bias with $n =$ 1000, $P =$ 500, $k =$ 1, MHR = 2 and censoring rate = 0}
\label{fig:n1000_p500_MHR2_censoring0}
\end{subfigure}\hfill
\begin{subfigure}{0.45\columnwidth}
\centering
\includegraphics[width=\textwidth]{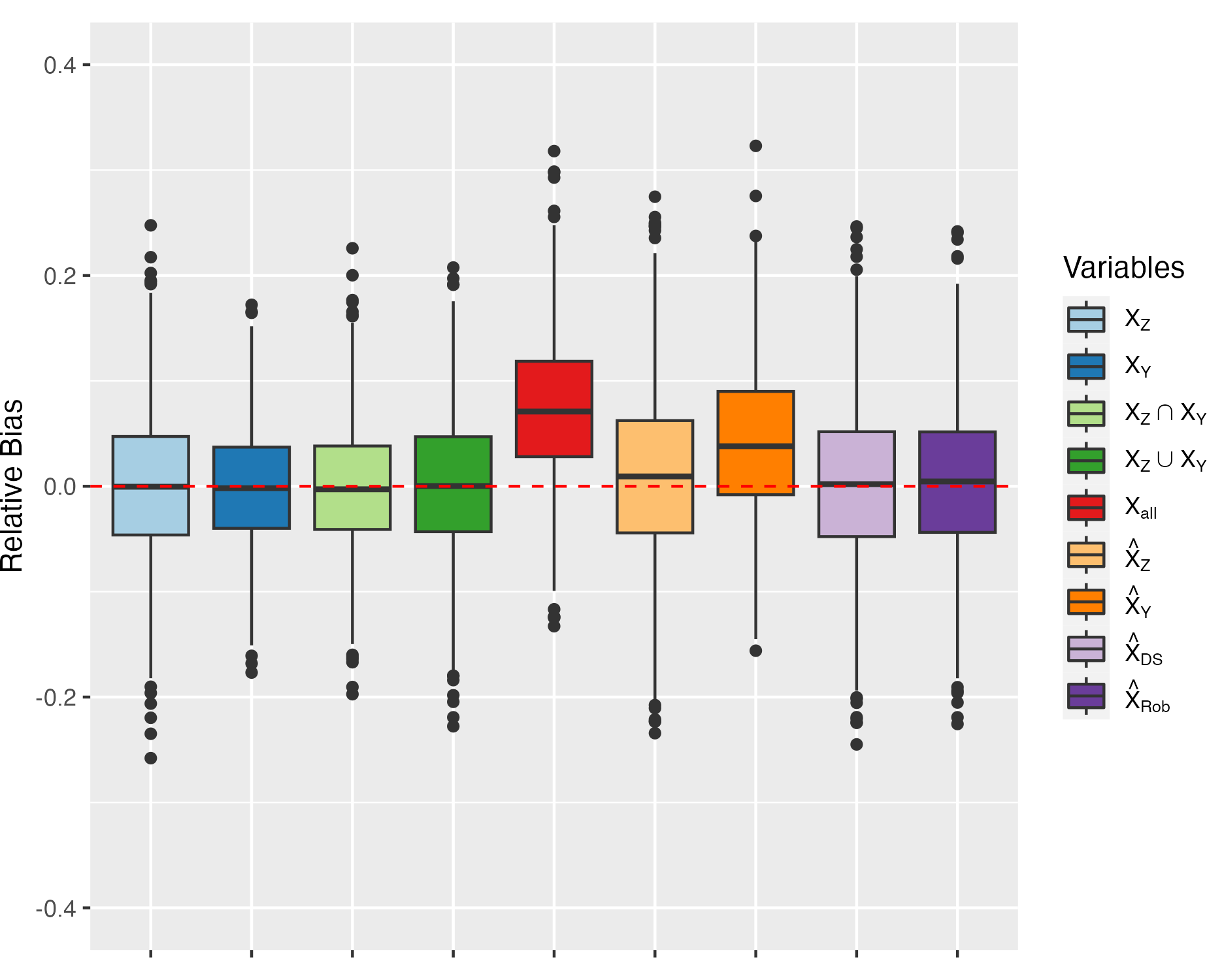}
\caption{Relative bias with $n =$ 1000, $P =$ 1000, $k =$ 1, MHR = 2 and censoring rate = 0}
\label{fig:Plot_n1000_p1000_MHR2_censoring0}
\end{subfigure}
\medskip
\begin{subfigure}{0.45\columnwidth}
\centering
\includegraphics[width=\textwidth]{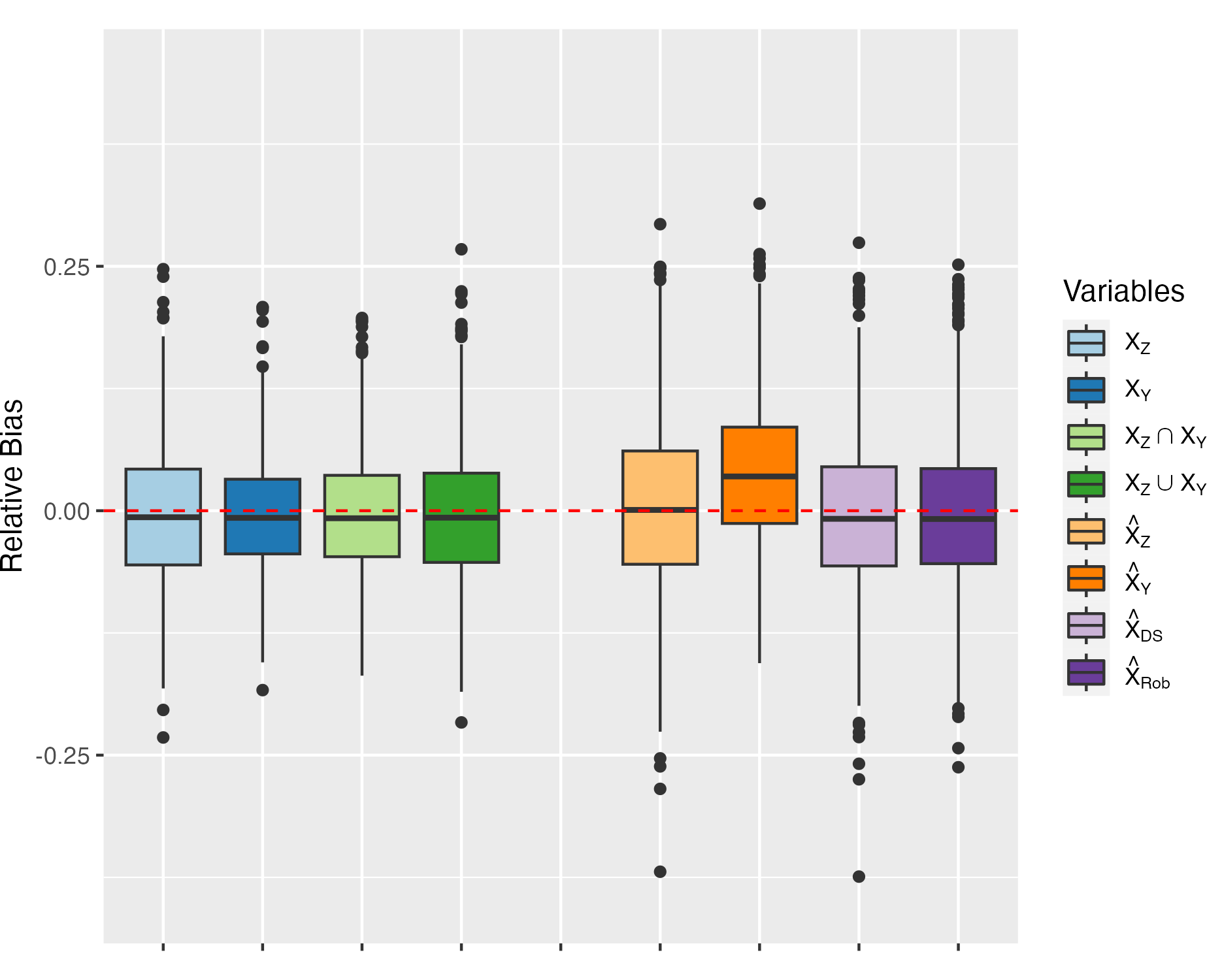}
\caption{Relative bias with $n =$ 1000, $P =$ 1500, $k =$ 1, MHR = 2 and censoring rate = 0}
\label{fig:Plot_n1000_p1500_MHR2_censoring0}
\end{subfigure}\hfill
\begin{subfigure}{0.45\columnwidth}
\centering
\includegraphics[width=\textwidth]{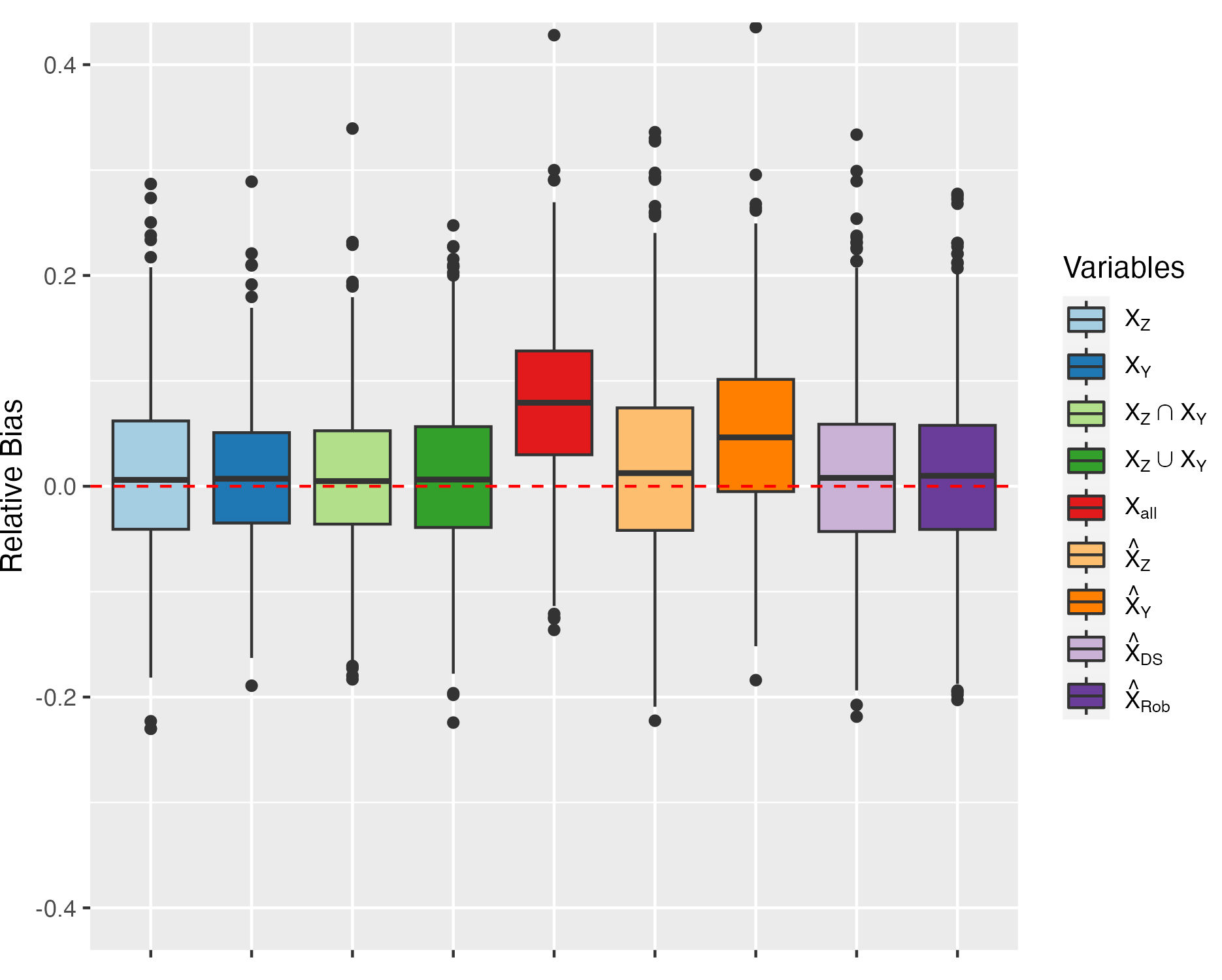}
\caption{Relative bias with $n =$ 1000, $P =$ 1000, $k =$ 1, MHR = 0.5 and censoring rate = 0}
\label{fig:Plot_n1000_p1000_MHR0.5_censoring0}
\end{subfigure}
\medskip
\begin{subfigure}{0.45\columnwidth}
\centering
\includegraphics[width=\textwidth]{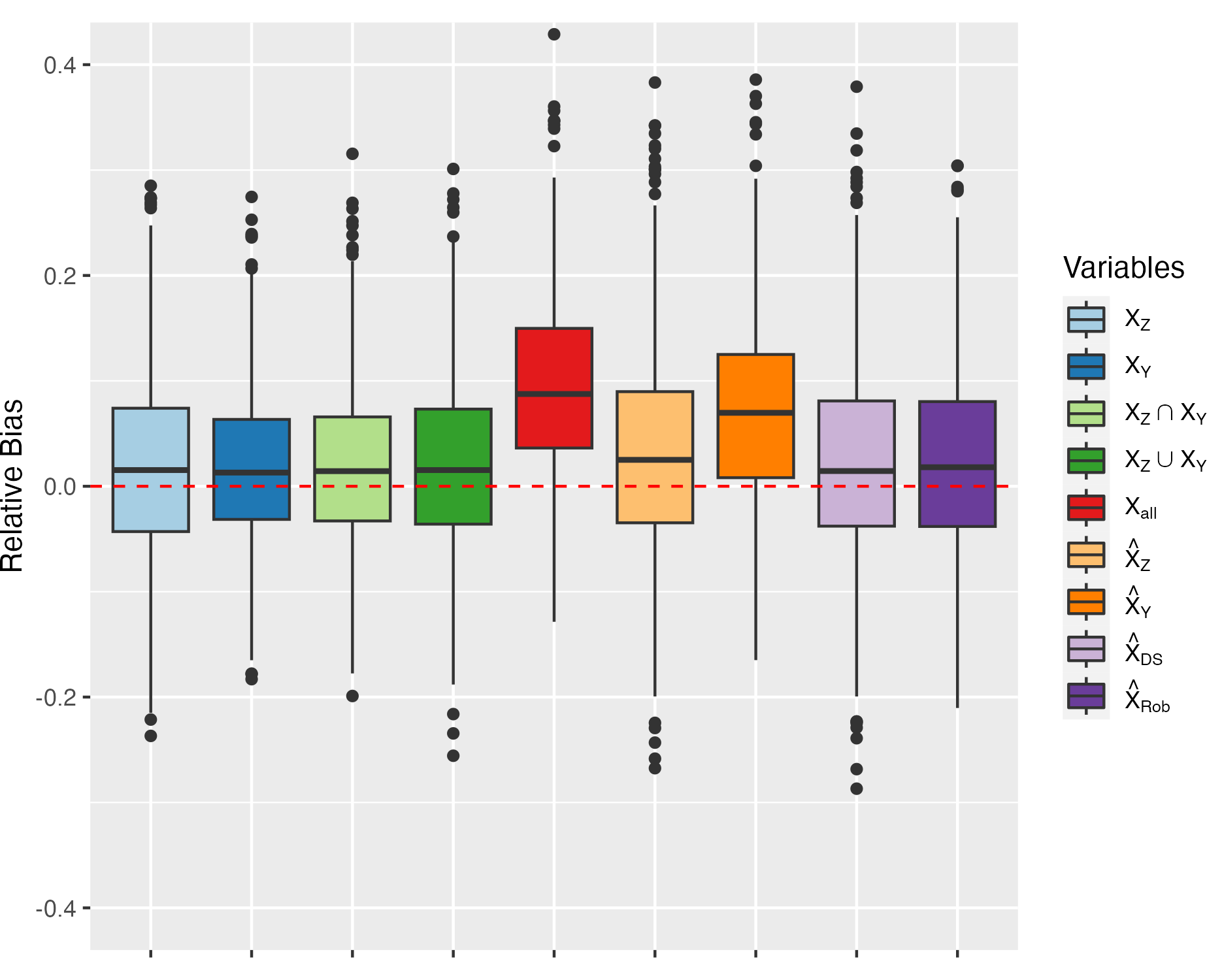}
\caption{Relative bias with $n =$ 1000, $P =$ 1000, $k =$ 1, MHR = 2 and censoring rate = 0.2}
\label{fig:Plot_n1000_p1000_MHR2_censoring0.2}
\end{subfigure}\hfill
\begin{subfigure}{0.45\columnwidth}
\centering
\includegraphics[width=\textwidth]{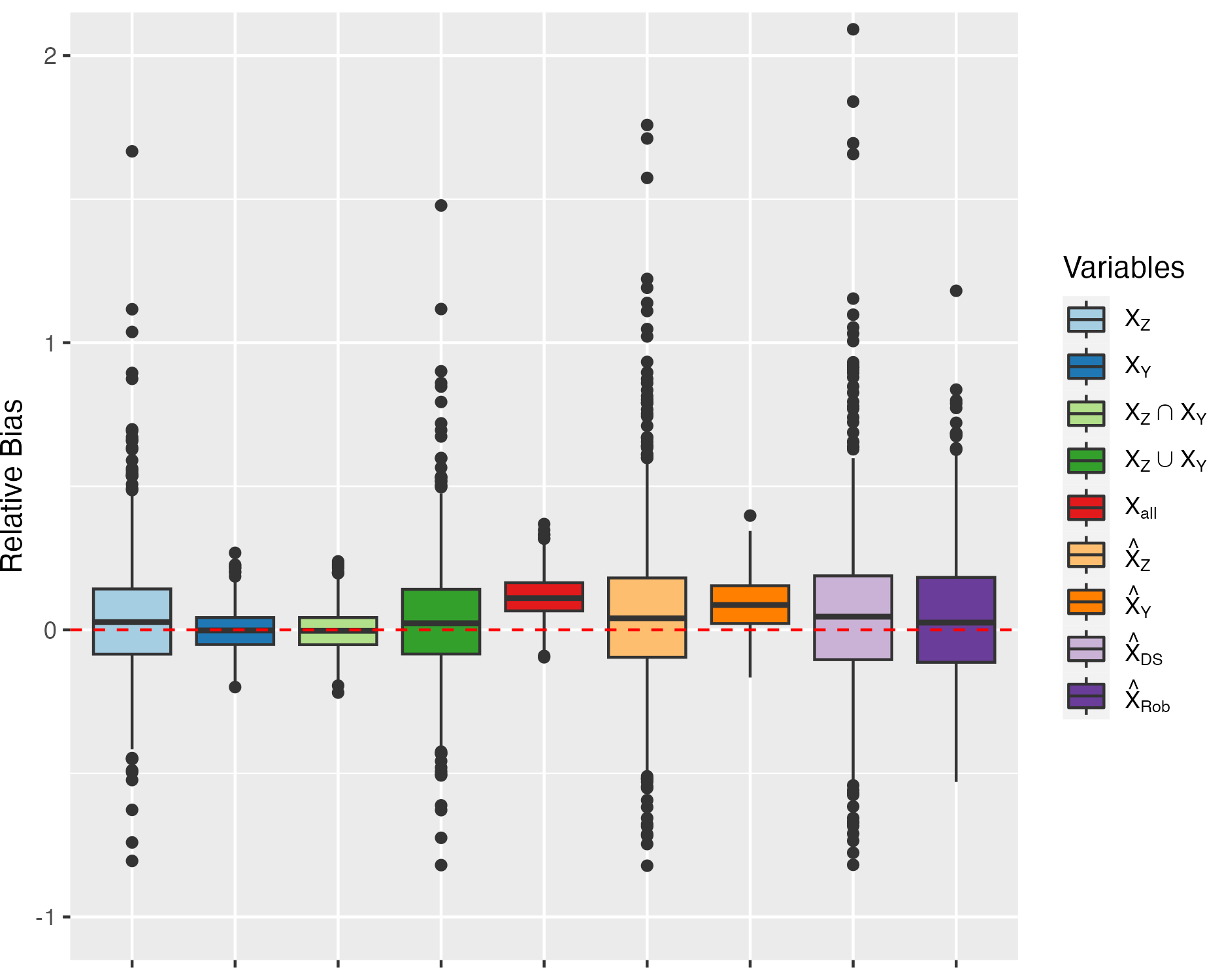}
\caption{Relative bias with $n =$ 1000, $P =$ 1000, $k =$ 3, MHR = 2 and censoring rate = 0.2}
\label{fig:Plot_n1000_p1000_MHR2_censoring0.2_k3}
\end{subfigure}

\caption{Relative bias of estimation of MHR for settings (1) - (6) for different variable selection strategies when creating weights. Results based on 1000 simulation replicates}
\label{fig:sumary_results}

\end{figure}

The summary of the simulation results can be seen in Figure \ref{fig:sumary_results} and further details are provided in Tables \ref{table:n1000_p500_MHR2_censoring0}-\ref{table:n500_p500_MHR2_censoring0_k3} in Appendix \ref{section:AppendixA}. Note that in the case where $n<p$, the results for $X_{all}$ are omitted due to the regression being infeasible.

From the results, we can see that once variable selection becomes relevant, even in the cases where $n>P$, using $X_{all}$ results in biased estimation of MHR. This is an issue similar to model misspecification, as the coefficients assigned to the covariates do not accurately reflect the true relationships between these variables and the treatment. Consequently, the resulting weights, when used in a weighted Cox regression, introduce bias in the estimation of MHR. 

Looking at the figures and the tables, it seems like the ideal scenario would be to generate weights based on a propensity score derived solely from $X_Y$. This set of covariates, which affects the outcome, outperforms even the true set of confounders ($X_Z \cap X_Y$) in terms of RMSE by reducing the variance without introducing bias. Similarly, the union set ($X_Z \cup X_Y$) shows better performance compared to the set of variables affecting only the treatment assignment ($X_Z$). This shows the importance of including the variables predicting outcome for reducing the variance when estimating MHR. It also indicates that while instrumental variables do increase the variance of the estimate, they do not impact the bias. 

In practice, none of these sets are known to the researcher. An improvement over the naive approach of using all variables would be to apply a lasso regression on the treatment as a response. This approach, aimed at selecting $X_Z$ (and thus is denoted $\hat{X}_Z$), is in the absence of censoring successful in estimating MHR without bias. Conversely, using lasso Cox regression in an attempt to reproduce the set $X_Y$ results in biased MHR estimation. This likely stems from  the fact that not all the confounders are included in the selected set $\hat{X}_Y$, as evidenced by the true positive rate of each set shown in Table \ref{table:variable_sel}. It is observed that $\hat{X}_Z$ and $\hat{X}_{DS}$ included the four main confounders $X_1, \dots, X_4$ in all repetitions, which is the necessary condition for unbiased estimation, whereas $\hat{X}_Y$ did not.

\begin{table}[ht]
\centering
\caption{Average true positive rate, and average and max F1 score for selection of the 4 confounders $X_1, \dots, X_4$ and the and average cardinality of each set in Scenario 6 for all N = 1000 simulated datasets }
\begin{tabular}{lrrrr}
  \hline
Set & Average True Positive Rate & Average F1 Score & Max F1 Score & Average Cardinality  \\ 
  \hline
$\hat{X}_Z$ & 4.00 & 0.39 & 0.73 & 19.9\\
 $\hat{X}_Y$ & 2.74 & 0.57 & 0.89 & 5.94\\
 $\hat{X}_{DS}$ & 4.00 & 0.34 & 0.62 & 23.0\\
 $\hat{X}_I$ & 2.74 & 0.78 & 1.00 & 2.79\\
   \hline
\end{tabular}
\label{table:variable_sel}
\end{table}

A more effective approach to mitigate this source of bias, similar to the strategy used in the non-time-to-event context, is the use of double selection. In our simulations, double selection did not yield results as optimal as those achieved by knowing any of the true sets ($X_Z$, $X_Y$ or their union or intersection), but it produced estimates with considerably lower variance compared to those obtained by selecting only the variables $\hat{X}_Z$, thus achieving a considerably better RMSE.

Our implementation of the multiply robust method in combination with double selection demonstrates promising results. In almost all simulations, its performance closely matches that of double selection alone, suggesting that double selection identifies the best set of variables most of the time. However, the multiply robust methods does result in slight improvements in variance and RMSE, indicating that there is some value in considering the estimated weights derived from the sets $\hat{X}_Z$, $\hat{X}_Y$ and $\hat{X}_I$. This improvement is particularly notable in more complex estimation scenarios, such as when the overlap is bad (exemplified in Scenario 6). This improvement stems from the fact that, although $\hat{X}_Y$ and/or $\hat{X}_I$ are often misspecified sets and generally underperform when compared with $\hat{X}_{DS}$, there are a few rare cases where $\hat{X}_Y$ and/or $\hat{X}_I$ lead to better weights for MHR estimation due to more precise variable selection. This can be seen in Table \ref{table:variable_sel} where the average F1 score for $\hat{X}_Y$ and $\hat{X}_I$ has higher values. Consequently, the combination of double selection with the multiply robust approach enables us to outperform $\hat{X}_{DS}$ in these instances, without falling into the dangers of model misspecification that causes bias due to the set $\hat{X}_Y$. The improvement observed with the use of the multiply robust estimator is also more notable in the scenarios with a smaller sample size, maintaining the same ratio between $n$ and $p$. Results for a lower sample size ($n = 500$) are presented in Tables \ref{table:n500_p250_MHR2_censoring0}-\ref{table:n500_p500_MHR2_censoring0_k3} in the Appendix, where it is evident that the improvements are generally more marked. 

In the scenarios with censoring, all sets result in biased MHR estimation, a phenomenon previously noted when estimating MHR under non-informative censoring \citep{wyss, fireman, hajage_censoring_bias}. However, this censoring-induced bias is also present when using the set of true confounders, and it does not appear to be exacerbated by using sets selected through variable selection methods, as compared to using known true variable sets.

\FloatBarrier

\section{Case Study}
\label{section:case_study}

To demonstrate the application of these covariate selection approaches in a real-world context, we examine the case of anticoagulant (Warfarin) prescription for patients following an initial stroke and subsequent hospital discharge.  The event of interest after hospital discharge is the occurrence of either a second stroke or death within the ten years following hospital discharge. 

Our initial sample consisted of 2884 patients born in Sweden who suffered their first ischemic stroke in 2006 while residing in Sweden. The data was sourced from the Swedish Stroke Register (Riksstroke) and linked to the Swedish Longitudinal Integrated Database for Health Insurance and Labour Market Studies (LISA; administered by Statistics Sweden) and the National Patient Register (NPR; administered by the National Board of Health and Welfare). The latter data sources provided information on additional covariates such as income, education, and comorbidity conditions. In the sample, 973 patients (33.75\%) were prescribed Wafarin upon discharge, and 2512 patients (87.13\%) experienced the event within the subsequent ten years. The covariates being considered are listed in Table \ref{table:real_data}. To assess the overlap in the sample, we estimated the propensity score with a logistic regression, including all the covariates in the table below. The overlap plot can be seen in Figure \ref{fig:Overlap_real_data}.

\begin{table}[ht]
\centering
 \caption{Baseline characteristics of treated and untreated subjects in the original sample for each covariate. }
\label{table:real_data}
\begin{tabular}{lrr}
 \hline
  Covariate &  \multicolumn{1}{c}{No anti-coagulant} & \multicolumn{1}{c}{Anti-coagulant} \\
  & \multicolumn{1}{c}{($n =$ 1910)} &  \multicolumn{1}{c}{($n =$ 973)} \\
 \hline
 \multicolumn{3}{l}{\textit{Demographic and background characteristics}} \\
 Age & $82.3 \pm 7.6$ & $75.3 \pm 8.6$  \\
 Female  & 1060 (55.5\%) & 417 (42.9\%)  \\
 Living alone  & 1164 (60.9\%) & 362 (37.2\%)  \\
 Living in an institution  & 216 (11.3\%) & 14 (1.4\%)  \\
 ADL dependency  & 227 (11.9\%) & 32 (3.3\%) \\
 \multicolumn{3}{l}{\textit{Socioeconomic and educational level}} \\
 Income & $1460 \pm 1564$ & $1712 \pm 1162$ \\
 Primary & 1218 (63.7\%) & 527 (54.2\%)  \\
 Secondary & 517 (27.0\%) & 300 (30.8\%)   \\
 University & 175 (9.2\%) & 146 (15.0\%)   \\
 \multicolumn{3}{l}{\textit{Level of consciousness at admission}} \\
 Alert  & 1587 (83.1\%) & 904 (92.9\%) \\
 Drowsy  & 275 (14.4\%) & 58 (6.0\%) \\
 Unconscious  & 48 (2.5\%) & 11 (1.1\%)  \\
  \multicolumn{3}{l}{\textit{Comorbid conditions}}\\
 Diabetes  & 410 (21.5\%) & 190 (19.5\%)  \\
 Smoking  & 151 (7.9\%) &  100 (10.3\%)  \\
 Hypertension medication  & 1239 (64.9\%) & 626 (64.3\%) \\
 Heart failure  & 544 (28.5\%) & 219 (22.5\%)  \\
 Ischemic heart disease  & 571  (29.9\%) &  250 (25.7\%) \\
 Dementia  &  129 (6.8\%) & 10 (1.0\%) \\
 Cancer in last three years  & 144 (7.5\%) &  67 (6.9\%) \\
 Valvular disease   & 133 (7.0\%) & 93 (9.6\%) \\
 Peripheral arterial disease   & 114 (6.0\%) &  49 (5.0\%) \\
 Venous thromboembolism   & 56 (2.9\%) & 31 (3.2\%) \\
 Intracerebral hemorrhage (I61)  & 37 (1.9\%) & 4 (0.4\%) \\
 Transient ischemic attack (TIA)  & 139 (7.3\%) & 65 (6.7\%) \\
 Other major bleeding   & 178 (9.3\%) & 49 (5.0\%) \\
 \hline
\end{tabular}
\end{table}

To mimic the conditions of the simulation study, we consider two scenarios: one using only the original covariates and and another with additional independent covariates, added to our sample until $n = P$. Half of these additional covariates follow a standard normal distribution, while the other half follow a Bernoulli(0.5) distribution. These added covariates are named X1 to X2884.

In Table \ref{table:real_data_sel_sets}, we present the selected sets $\hat{X}_Z$, $\hat{X}_Y$, $\hat{X}_{DS}$, and $\hat{X}_I$. The estimated MHR resulting from these different sets, along with the application of the multiply robust approach, are detailed in Table \ref{table:real_data_MHR}.

\begin{table}[ht]
\centering
\caption{Covariate sets selected by lasso procedure on treatment and outcome, their union and intersection.}
\begin{tabular}{lc}

 \hline
 Set & Covariates in the Set \\
  \hline
 $\hat{X}_Z$ & Age, Drowsy, Living in an institution, Living Alone, ADL dependency, Dementia, X1084  \\
 $\hat{X}_Y$ & Age, Heart Failure, Living in an institution, ADL dependency \\
 $\hat{X}_{DS}$ & Age, Drowsy, Living in an institution, Living Alone, ADL dependency, Dementia, X1084, Heart Failure \\
 $\hat{X}_I$ & Age, Living in an institution, ADL-dependency \\

 \hline
\end{tabular}
\label{table:real_data_sel_sets}
\end{table}

\begin{table}[ht]
\centering
\caption{MHR estimations from the propensity score and weights generated by different covariate sets.}
\begin{tabular}{lccc}

 \hline
 Covariate Set & Estimated MHR & CIL & CIU\\
  \hline
 $\hat{X}_Z$ &   0.682 & 0.617 & 0.753\\
 $\hat{X}_Y$ &  0.668 & 0.606 & 0.736\\
 $\hat{X}_{DS}$ & 0.681 & 0.616 & 0.752\\
 $\hat{X}_{Rob}$ & 0.677 & 0.611 & 0.749\\

 \hline
\end{tabular}
\label{table:real_data_MHR}
\end{table}

We can see that in our case study, which includes a reasonably large number of individuals and good overlap, the estimated MHRs are in general quite similar across all sets and methods. However, it should be noted that both $\hat{X}_Z$ and $\hat{X}_{DS}$ included an artificially generated covariate, X1084, which is definitively not a confounder. On the other hand, $\hat{X}_Y$ includes a significantly smaller number of covariates but includes 'Heart Failure', a covariate that is not included in $\hat{X}_Z$. The MHR value estimated by $\hat{X}_{Rob}$ is between those of $\hat{X}_{DS}$ and $\hat{X}_Y$, but closer to $\hat{X}_{DS}$, a result in line with what was observed in our simulations.

\FloatBarrier
\section{Discussion}
\label{section:discussion}

While the issue of model selection and bias due to model misspecification in the context of MHR estimation has been previously been explored \citep{austin_fullmatching_misspecification, multi_robust}, past studies primarily focused on scenarios with relatively few variables. In contrast, the high-dimensional scenario, has received considerable attention within the causal inference framework \citep{causal_inf_damour, causal_inf_avagyan, causal_inf_hill, causal_inf_gao, causal_inf_antonelli, selection_de_luna, selection_haggstrom, selection_persson} and it is also a well-known challenge in time-to-event studies and Cox regression for conditional hazard ratios \citep{survival_high_dim_witten, survival_high_dim_guide_salerno, survival_high_dim_Lasso_application_xu, survival_high_dim_waldron}. Taking these points into account, this paper explores the problem of variable and model selection for estimating MHR in high-dimensional datasets through the use of Monte Carlo simulations, employing methods developed for model selection in the time-to-event setting and for variable selection within the causal inference framework. 

Our simulation results show that MHR estimation is, as expected, analogous to results in the causal inference literature when using the true sets $X_Y$,  $X_Z$, $X_{Z} \cap X_Y$ and $X_{Z} \cup X_Y$. We reproduced variable selection results reported by \citet{causal_inf_brookhart}, in which using covariates solely related to the outcome for propensity score and weight estimation leads to smaller variance and improved RMSE when estimating MHR. Furthermore, our simulations support the claim that including instrumental variables increases variance, as previously reported in the literature. 

However, when dealing with sets selected via lasso, $\hat{X}_Y$ does induce some bias due to not including all confounders. A viable solution to this issue is the application of double selection which, in our simulation scenarios, generally performed well in terms of RSME. This is a relevant finding since double selection is a method that can be easily and quickly implemented in practical research while yielding estimates that closely approximate those of ideal scenarios.

Our results also demonstrate that double selection can be improved upon by the use of a multiply robust approach including the sets of covariates selected by lasso as well as their union and intersection. This approach does not sacrifice any performance in relation to bias, while simultaneously reducing the variance, leading to improved performance in terms of RMSE. These results and behaviors were shown across a diverse range of scenarios, including cases with true MHR above and below 1, varying ratios of sample size to number of covariates ($n<p$, $n>p$, $n=p$), involving censoring, and with bad and good overlap. Notably, this improvement is particularly pronounced in scenarios with bad overlap, where not having access to the true sets $X_Y$ and $X_Z$ greatly diminishes performance, even for double selection. Additionally, our results indicate that the differences in performance between different estimated covariate sets decreases as the sample size increases, even when the number of covariates also increases.

However, an automated double selection method, like the one explored in this paper, has its limitations and should be utilized in conjunction with researcher expertise and experience. For instance, the inclusion of endogenous variables is not something that is solved by double selection \citep{good_bad_controls}, and including such variables would lead to biased estimates \citep{DML_criticism}. Although this paper does not explore this aspect in the context of MHR estimation, researchers should carefully consider which variables ought to be subjected to double selection methods.

In the case study, we show conclusions similar to the simulations even when working with real data. With a larger dataset and strong overlap, the estimated MHRs are similar regardless of which of the covariate sets that was used. We observed that selecting for $X_Y$ yielded a smaller set of covariates compared to selecting for $X_Z$, although the latter included a covariate known to be irrelevant. $\hat{X}_Y$ also includes covariates that were not found in $\hat{X}_Z$. The results also indicate that the use of anti-coagulants post-discharge does have a protective effect against a second stroke and increases the survival of the patient. This finding aligns with other studies that have shown the benefits of such prescriptions \cite{AC_Germany, AC_Klijn, AC_Xian, AC_Sweden}.

In conclusion, our simulation results support the use of double selection for selecting covariates when estimating MHR by weighted Cox regression, as this method alone yielded good results. Further improvements might be achieved by combining double selection with the use of the multiply robust approach. For future research in the time-to-event context, we recommend exploreing other variable selection methods besides lasso to assess if further improvements are possible. 

\section*{Acknowledgements}
The authors are grateful to Professor Xavier de Luna for helpful and constructive comments. This work was supported by the Swedish Research Council 377 (Dnr: 2018–01610).

\section*{Ethical considerations}

\noindent 
Statistical method development for fair comparisons of stroke care and outcome was part of the EqualStroke-project, approved by the Ethical Review Board in Umeå (Dnr: 2012-321-31M, 2014-76-32M). Patients and next of kin are informed about the registration and aim of the Riksstroke-register and their right to decline participation (opt-out consent).

\nocite{*}

\printbibliography[category=cited]


\FloatBarrier

\newpage
\appendix
\counterwithin{figure}{section}
\counterwithin{table}{section}

\section{Appendix A}
\label{section:AppendixA}

\begin{figure}[htp]
\centering

\begin{subfigure}{0.45\columnwidth}
\centering
\includegraphics[width=\textwidth]{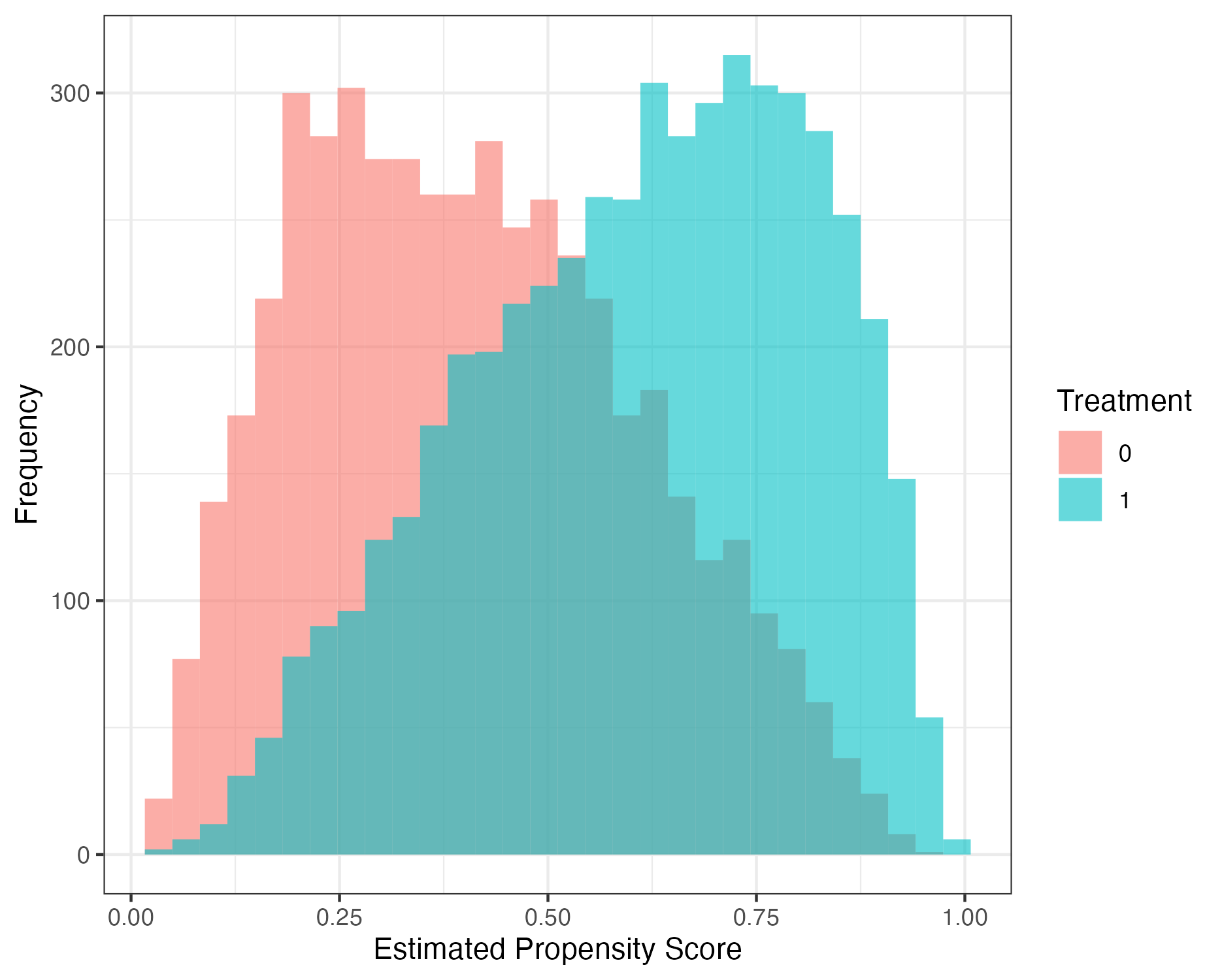}
\caption{Overlap of a generated dataset when $k =$ 1}
\label{fig:example_k_1}
\end{subfigure}\hfill
\begin{subfigure}{0.45\columnwidth}
\centering
\includegraphics[width=\textwidth]{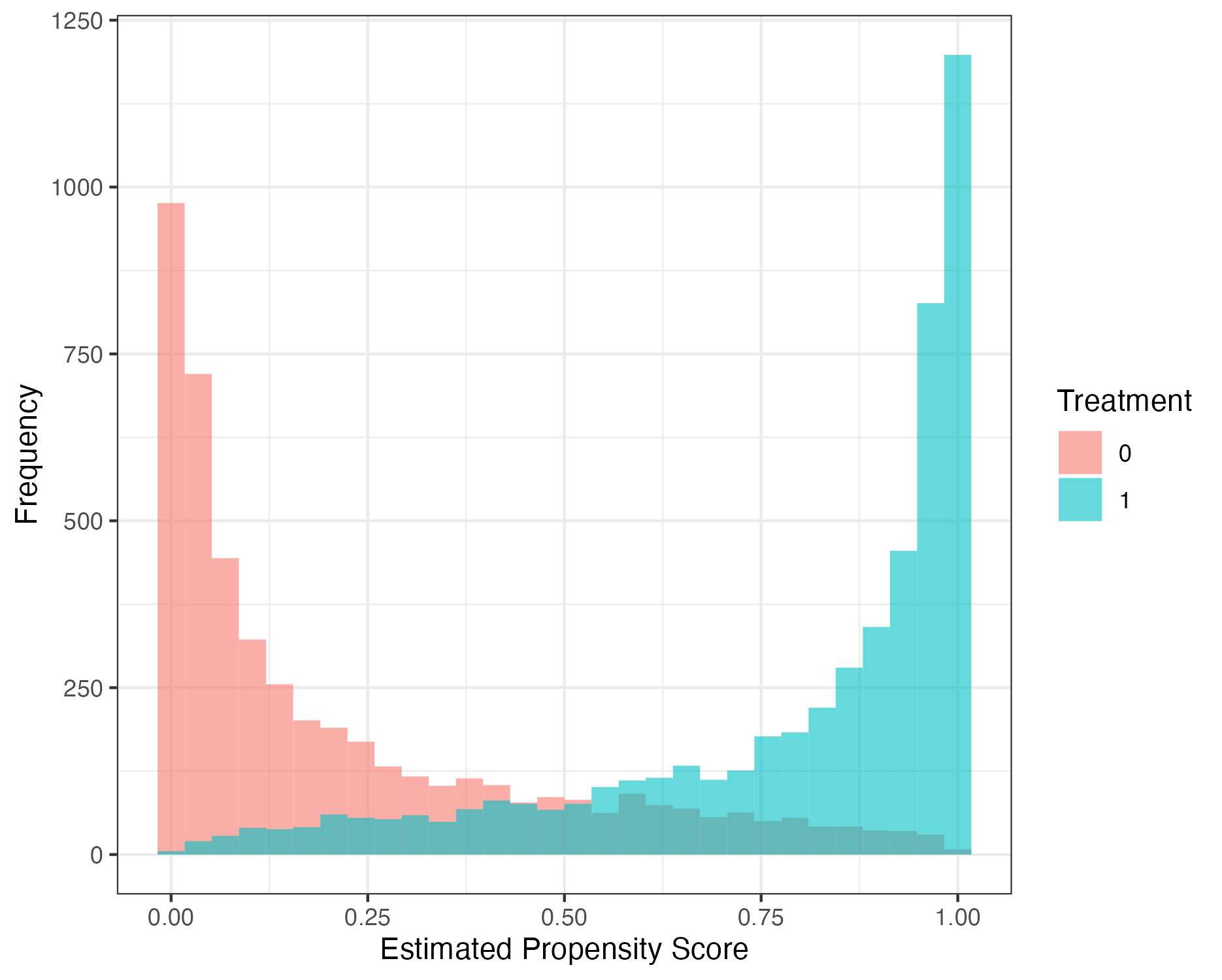}
\caption{Overlap of a generated dataset when $k =$ 3}
\label{fig:example_k_3}
\end{subfigure}
\medskip

\caption{Overlap example dataset with $k =$ 1 and $k =$ 3}
\label{fig:k_example}

\end{figure}

\begin{table}[ht]
\centering
\caption{Simulation results with $n =$ 1000, $P =$ 500, $k =$ 1, MHR = 2 and no censoring rate. Results of 1000 simulation replicates for estimators of MHR for $X_{Cf}$, $X_Z$, $X_{all}$, $\hat{X}_Z$, $\hat{X}_{DS}$ and $\hat{X}_{Rob}$. Bias, Monte Carlo standard deviation (SD), root mean-squared error (RMSE), and empirical coverage probability of 95\% confidence intervals (Coverage)} 
\begin{tabular}{lrrrr}
  \hline
 & Bias & SD & RMSE & Coverage \\ 
  \hline
$X_Z$ & 0.003 & 0.144 & 0.144 & 0.970 \\ 
  $X_Y$ & -0.004 & 0.113 & 0.113 & 0.985 \\ 
  $X_Z \cap X_Y$ & -0.002 & 0.121 & 0.121 & 0.982 \\ 
  $X_Z \cup X_Y$ & 0.002 & 0.136 & 0.136 & 0.984 \\ 
  $X_{all}$ & 0.146 & 0.142 & 0.204 & 0.854 \\ 
  $\hat{X}_Z$ & 0.016 & 0.161 & 0.162 & 0.964 \\ 
  $\hat{X}_Y$ & 0.068 & 0.138 & 0.154 & 0.938 \\ 
  $\hat{X}_{DS}$ & 0.004 & 0.149 & 0.149 & 0.978 \\ 
  $\hat{X}_{Rob}$ & 0.003 & 0.145 & 0.145 & 0.981 \\ 
   \hline
\end{tabular}
\label{table:n1000_p500_MHR2_censoring0}
\end{table}

\begin{table}[ht]
\centering
\caption{Simulation results with $n =$ 1000, $P =$ 1000, $k =$ 1, MHR = 2 and no censoring rate. Results of 1000 simulation replicates for estimators of MHR for $X_{Cf}$, $X_Z$, $X_{all}$, $\hat{X}_Z$, $\hat{X}_{DS}$ and $\hat{X}_{Rob}$. Bias, Monte Carlo standard deviation (SD), root mean-squared error (RMSE), and empirical coverage probability of 95\% confidence intervals (Coverage)} 
\begin{tabular}{lrrrr}
  \hline
 & Bias & SD & RMSE & Coverage \\ 
  \hline
$X_Z$ & 0.002 & 0.142 & 0.142 & 0.981 \\ 
  $X_Y$ & -0.001 & 0.112 & 0.112 & 0.986 \\ 
  $X_Z \cap X_Y$ & -0.000 & 0.121 & 0.121 & 0.977 \\ 
  $X_Z \cup X_Y$ & 0.003 & 0.135 & 0.135 & 0.987 \\ 
  $X_{all}$ & 0.146 & 0.136 & 0.199 & 0.852 \\ 
  $\hat{X}_Z$ & 0.021 & 0.169 & 0.170 & 0.958 \\ 
  $\hat{X}_Y$ & 0.084 & 0.141 & 0.163 & 0.934 \\ 
  $\hat{X}_{DS}$ & 0.004 & 0.153 & 0.153 & 0.983 \\ 
  $\hat{X}_{Rob}$ & 0.007 & 0.145 & 0.145 & 0.983 \\ 
   \hline
\end{tabular}
\label{table:n1000_p1000_MHR2_censoring0}
\end{table}

\begin{table}[ht]
\centering
\caption{Simulation results with $n =$ 1000, $P =$ 1500, $k =$ 1, MHR = 2 and no censoring rate. Results of 1000 simulation replicates for estimators of MHR for $X_{Cf}$, $X_Z$, $X_{all}$, $\hat{X}_Z$, $\hat{X}_{DS}$ and $\hat{X}_{Rob}$. Bias, Monte Carlo standard deviation (SD), root mean-squared error (RMSE), and empirical coverage probability of 95\% confidence intervals (Coverage)} 
\begin{tabular}{lrrrr}
  \hline
 & Bias & SD & RMSE & Coverage \\ 
  \hline
$X_Z$ & -0.009 & 0.147 & 0.148 & 0.976 \\ 
  $X_Y$ & -0.010 & 0.119 & 0.119 & 0.982 \\ 
  $X_Z \cap X_Y$ & -0.010 & 0.127 & 0.127 & 0.970 \\ 
  $X_Z \cup X_Y$ & -0.008 & 0.139 & 0.139 & 0.978 \\ 
  $X_{all}$ &  &  &  &  \\ 
  $\hat{X}_Z$ & 0.011 & 0.174 & 0.174 & 0.954 \\ 
  $\hat{X}_Y$ & 0.078 & 0.146 & 0.166 & 0.923 \\ 
  $\hat{X}_{DS}$ & -0.007 & 0.157 & 0.157 & 0.974 \\ 
  $\hat{X}_{Rob}$ & -0.007 & 0.153 & 0.153 & 0.976 \\ 
   \hline
\end{tabular}
\label{table:n1000_p1500_MHR2_censoring0}
\end{table}

\begin{table}[ht]
\centering
\caption{Simulation results with $n =$ 1000, $P =$ 1000, $k =$ 1, MHR = 0.5 and no censoring rate. Results of 1000 simulation replicates for estimators of MHR for $X_{Cf}$, $X_Z$, $X_{all}$, $\hat{X}_Z$, $\hat{X}_{DS}$ and $\hat{X}_{Rob}$. Bias, Monte Carlo standard deviation (SD), root mean-squared error (RMSE), and empirical coverage probability of 95\% confidence intervals (Coverage)} 
\begin{tabular}{lrrrr}
  \hline
 & Bias & SD & RMSE & Coverage \\ 
  \hline
$X_Z$ & 0.005 & 0.037 & 0.038 & 0.958 \\ 
  $X_Y$ & 0.004 & 0.032 & 0.033 & 0.968 \\ 
  $X_Z \cap X_Y$ & 0.005 & 0.034 & 0.034 & 0.958 \\ 
  $X_Z \cup X_Y$ & 0.005 & 0.036 & 0.036 & 0.965 \\ 
  $X_{all}$ & 0.040 & 0.037 & 0.055 & 0.792 \\ 
  $\hat{X}_Z$ & 0.009 & 0.043 & 0.044 & 0.942 \\ 
  $\hat{X}_Y$ & 0.025 & 0.039 & 0.046 & 0.862 \\ 
  $\hat{X}_{DS}$ & 0.005 & 0.039 & 0.040 & 0.963 \\ 
  $\hat{X}_{Rob}$ & 0.005 & 0.038 & 0.038 & 0.960 \\ 
   \hline
\end{tabular}
\label{table:n1000_p1000_MHR0.5_censoring0}
\end{table}

\begin{table}[ht]
\centering
\caption{Simulation results with $n =$ 1000, $P =$ 1000, $k =$ 1, MHR = 2 and censoring rate of 20\%. Results of 1000 simulation replicates for estimators of MHR for $X_{Cf}$, $X_Z$, $X_{all}$, $\hat{X}_Z$, $\hat{X}_{DS}$ and $\hat{X}_{Rob}$. Bias, Monte Carlo standard deviation (SD), root mean-squared error (RMSE), and empirical coverage probability of 95\% confidence intervals (Coverage)} 
\begin{tabular}{lrrrr}
  \hline
 & Bias & SD & RMSE & Coverage \\ 
  \hline
$X_Z$ & 0.038 & 0.172 & 0.176 & 0.967 \\ 
  $X_Y$ & 0.034 & 0.140 & 0.145 & 0.973 \\ 
  $X_Z \cap X_Y$ & 0.035 & 0.149 & 0.153 & 0.962 \\ 
  $X_Z \cup X_Y$ & 0.038 & 0.164 & 0.168 & 0.981 \\ 
  $X_{all}$ & 0.186 & 0.163 & 0.247 & 0.790 \\ 
  $\hat{X}_Z$ & 0.059 & 0.197 & 0.206 & 0.935 \\ 
  $\hat{X}_Y$ & 0.138 & 0.172 & 0.220 & 0.863 \\ 
  $\hat{X}_{DS}$ & 0.043 & 0.184 & 0.189 & 0.964 \\ 
  $\hat{X}_{Rob}$ & 0.045 & 0.174 & 0.179 & 0.977 \\ 
   \hline
\end{tabular}
\label{table:n1000_p1000_MHR2_censoring0.2}
\end{table}

\begin{table}[ht]
\centering
\caption{Simulation results with $n =$ 1000, $P =$ 1000, $k =$ 3, MHR = 2 and no censoring rate. Results of 1000 simulation replicates for estimators of MHR for $X_{Cf}$, $X_Z$, $X_{all}$, $\hat{X}_Z$, $\hat{X}_{DS}$ and $\hat{X}_{Rob}$. Bias, Monte Carlo standard deviation (SD), root mean-squared error (RMSE), and empirical coverage probability of 95\% confidence intervals (Coverage)} 
\begin{tabular}{lrrrr}
  \hline
 & Bias & SD & RMSE & Coverage \\ 
  \hline
$X_Z$ & 0.070 & 0.419 & 0.425 & 0.919 \\ 
  $X_Y$ & -0.005 & 0.140 & 0.140 & 0.974 \\ 
  $X_Z \cap X_Y$ & -0.002 & 0.147 & 0.147 & 0.973 \\ 
  $X_Z \cup X_Y$ & 0.067 & 0.412 & 0.418 & 0.921 \\ 
  $X_{all}$ & 0.232 & 0.147 & 0.274 & 0.664 \\ 
  $\hat{X}_Z$ & 0.142 & 0.722 & 0.736 & 0.873 \\ 
  $\hat{X}_Y$ & 0.174 & 0.189 & 0.257 & 0.743 \\ 
  $\hat{X}_{DS}$ & 0.140 & 0.776 & 0.788 & 0.886 \\ 
  $\hat{X}_{Rob}$ & 0.089 & 0.432 & 0.441 & 0.929 \\ 
   \hline
\end{tabular}
\label{table:n1000_p1000_MHR2_censoring0_k3}
\end{table}

\FloatBarrier

\begin{table}[ht]
\caption{Simulation results with $n =$ 500, $P =$ 250, $k =$ 1, MHR = 2 and no censoring rate. Results of 1000 simulation replicates for estimators of MHR for $X_{Cf}$, $X_Z$, $X_{all}$, $\hat{X}_Z$, $\hat{X}_{DS}$ and $\hat{X}_{Rob}$. Bias, Monte Carlo standard deviation (SD), root mean-squared error (RMSE), and empirical coverage probability of 95\% confidence intervals (Coverage)} 
\centering
\begin{tabular}{lrrrr}
  \hline
 & Bias & SD & RMSE & Coverage \\ 
  \hline
$X_Z$ & 0.006 & 0.204 & 0.204 & 0.969 \\ 
  $X_Y$ & 0.000 & 0.169 & 0.169 & 0.978 \\ 
  $X_Z \cap X_Y$ & -0.001 & 0.180 & 0.180 & 0.975 \\ 
  $X_Z \cup X_Y$ & 0.007 & 0.193 & 0.193 & 0.981 \\ 
  $X_{all}$ & 0.344 & 6.174 & 6.184 & 0.908 \\ 
  $\hat{X}_Z$ & 0.035 & 0.241 & 0.244 & 0.954 \\ 
  $\hat{X}_Y$ & 0.099 & 0.194 & 0.218 & 0.948 \\ 
  $\hat{X}_{DS}$ & 0.019 & 0.225 & 0.225 & 0.971 \\ 
  $\hat{X}_{Rob}$ & 0.020 & 0.219 & 0.220 & 0.974 \\ 
   \hline
\end{tabular}
\label{table:n500_p250_MHR2_censoring0}
\end{table}

\begin{table}[ht]
\caption{Simulation results with $n =$ 500, $P =$ 500, $k =$ 1, MHR = 2 and no censoring rate. Results of 1000 simulation replicates for estimators of MHR for $X_{Cf}$, $X_Z$, $X_{all}$, $\hat{X}_Z$, $\hat{X}_{DS}$ and $\hat{X}_{Rob}$. Bias, Monte Carlo standard deviation (SD), root mean-squared error (RMSE), and empirical coverage probability of 95\% confidence intervals (Coverage)} 
\centering
\begin{tabular}{lrrrr}
  \hline
 & Bias & SD & RMSE & Coverage \\ 
  \hline
$X_Z$ & 0.007 & 0.213 & 0.213 & 0.964 \\ 
  $X_Y$ & -0.004 & 0.167 & 0.168 & 0.983 \\ 
  $X_Z \cap X_Y$ & -0.002 & 0.181 & 0.181 & 0.966 \\ 
  $X_Z \cup X_Y$ & 0.004 & 0.198 & 0.198 & 0.976 \\ 
  $X_{all}$ & 0.151 & 0.201 & 0.251 & 0.908 \\ 
  $\hat{X}_Z$ & 0.043 & 0.252 & 0.256 & 0.946 \\ 
  $\hat{X}_Y$ & 0.111 & 0.192 & 0.222 & 0.949 \\ 
  $\hat{X}_{DS}$ & 0.023 & 0.245 & 0.246 & 0.966 \\ 
  $\hat{X}_{Rob}$ & 0.025 & 0.227 & 0.228 & 0.966 \\ 
   \hline
\end{tabular}
\label{table:n500_p500_MHR2_censoring0}
\end{table}

\begin{table}[ht]
\caption{Simulation results with $n =$ 500, $P =$ 750, $k =$ 1, MHR = 2 and no censoring rate. Results of 1000 simulation replicates for estimators of MHR for $X_{Cf}$, $X_Z$, $X_{all}$, $\hat{X}_Z$, $\hat{X}_{DS}$ and $\hat{X}_{Rob}$. Bias, Monte Carlo standard deviation (SD), root mean-squared error (RMSE), and empirical coverage probability of 95\% confidence intervals (Coverage)} 
\centering
\begin{tabular}{lrrrr}
  \hline
 & Bias & SD & RMSE & Coverage \\ 
  \hline
$X_Z$ & -0.000 & 0.201 & 0.201 & 0.981 \\ 
  $X_Y$ & -0.013 & 0.161 & 0.161 & 0.989 \\ 
  $X_Z \cap X_Y$ & -0.009 & 0.174 & 0.174 & 0.977 \\ 
  $X_Z \cup X_Y$ & -0.004 & 0.189 & 0.189 & 0.981 \\ 
  $X_{all}$ &  &  &  &  \\ 
  $\hat{X}_Z$ & 0.041 & 0.269 & 0.272 & 0.951 \\ 
  $\hat{X}_Y$ & 0.112 & 0.198 & 0.228 & 0.944 \\ 
  $\hat{X}_{DS}$ & 0.033 & 0.330 & 0.332 & 0.966 \\ 
  $\hat{X}_{Rob}$ & 0.024 & 0.233 & 0.234 & 0.967 \\ 
   \hline
\end{tabular}
\label{table:n500_p750_MHR2_censoring0}
\end{table}

\begin{table}[ht]
\caption{Simulation results with $n =$ 500, $P =$ 500, $k =$ 1, MHR = 0.5 and no censoring rate. Results of 1000 simulation replicates for estimators of MHR for $X_{Cf}$, $X_Z$, $X_{all}$, $\hat{X}_Z$, $\hat{X}_{DS}$ and $\hat{X}_{Rob}$. Bias, Monte Carlo standard deviation (SD), root mean-squared error (RMSE), and empirical coverage probability of 95\% confidence intervals (Coverage)} 
\centering
\begin{tabular}{lrrrr}
  \hline
 & Bias & SD & RMSE & Coverage \\ 
  \hline
$X_Z$ & 0.007 & 0.055 & 0.055 & 0.949 \\ 
  $X_Y$ & 0.005 & 0.046 & 0.046 & 0.966 \\ 
  $X_Z \cap X_Y$ & 0.005 & 0.048 & 0.049 & 0.956 \\ 
  $X_Z \cup X_Y$ & 0.006 & 0.052 & 0.052 & 0.966 \\ 
  $X_{all}$ & 0.044 & 0.054 & 0.069 & 0.853 \\ 
  $\hat{X}_Z$ & 0.016 & 0.065 & 0.067 & 0.926 \\ 
  $\hat{X}_Y$ & 0.034 & 0.053 & 0.063 & 0.888 \\ 
  $\hat{X}_{DS}$ & 0.010 & 0.064 & 0.065 & 0.949 \\ 
  $\hat{X}_{Rob}$ & 0.011 & 0.060 & 0.061 & 0.955 \\ 
   \hline
\end{tabular}
\label{table:n500_p500_MHR0.5_censoring0}
\end{table}

\begin{table}[ht]
\caption{Simulation results with $n =$ 500, $P =$ 500, $k =$ 1, MHR = 2 and censoring rate of 20\%. Results of 1000 simulation replicates for estimators of MHR for $X_{Cf}$, $X_Z$, $X_{all}$, $\hat{X}_Z$, $\hat{X}_{DS}$ and $\hat{X}_{Rob}$. Bias, Monte Carlo standard deviation (SD), root mean-squared error (RMSE), and empirical coverage probability of 95\% confidence intervals (Coverage)} 
\centering
\begin{tabular}{lrrrr}
  \hline
 & Bias & SD & RMSE & Coverage \\ 
  \hline
$X_Z$ & 0.038 & 0.257 & 0.260 & 0.943 \\ 
  $X_Y$ & 0.029 & 0.203 & 0.205 & 0.969 \\ 
  $X_Z \cap X_Y$ & 0.031 & 0.217 & 0.220 & 0.956 \\ 
  $X_Z \cup X_Y$ & 0.035 & 0.243 & 0.246 & 0.958 \\ 
  $X_{all}$ & 0.190 & 0.237 & 0.304 & 0.874 \\ 
  $\hat{X}_Z$ & 0.071 & 0.304 & 0.312 & 0.936 \\ 
  $\hat{X}_Y$ & 0.153 & 0.238 & 0.283 & 0.908 \\ 
  $\hat{X}_{DS}$ & 0.054 & 0.304 & 0.309 & 0.949 \\ 
  $\hat{X}_{Rob}$ & 0.058 & 0.277 & 0.283 & 0.950 \\ 
   \hline
\end{tabular}
\label{table:n500_p500_MHR2_censoring0.2}
\end{table}

\begin{table}[ht]
\caption{Simulation results with $n =$ 500, $P =$ 500, $k =$ 3, MHR = 2 and no censoring rate. Results of 1000 simulation replicates for estimators of MHR for $X_{Cf}$, $X_Z$, $X_{all}$, $\hat{X}_Z$, $\hat{X}_{DS}$ and $\hat{X}_{Rob}$. Bias, Monte Carlo standard deviation (SD), root mean-squared error (RMSE), and empirical coverage probability of 95\% confidence intervals (Coverage)} 
\centering
\begin{tabular}{lrrrr}
  \hline
 & Bias & SD & RMSE & Coverage \\ 
  \hline
$X_Z$ & 0.111 & 0.613 & 0.623 & 0.910 \\ 
  $X_Y$ & 0.004 & 0.198 & 0.198 & 0.983 \\ 
  $X_Z \cap X_Y$ & 0.006 & 0.208 & 0.208 & 0.978 \\ 
  $X_Z \cup X_Y$ & 0.120 & 0.685 & 0.695 & 0.915 \\ 
  $X_{all}$ & 0.245 & 0.210 & 0.323 & 0.810 \\ 
  $\hat{X}_Z$ & 0.212 & 1.183 & 1.202 & 0.869 \\ 
  $\hat{X}_Y$ & 0.221 & 0.241 & 0.327 & 0.829 \\ 
  $\hat{X}_{DS}$ & 0.236 & 1.195 & 1.218 & 0.861 \\ 
  $\hat{X}_{Rob}$ & 0.155 & 0.706 & 0.723 & 0.899 \\ 
   \hline
\end{tabular}
\label{table:n500_p500_MHR2_censoring0_k3}
\end{table}

\FloatBarrier

\begin{figure}[htp]
\centering
\includegraphics[width=\textwidth]{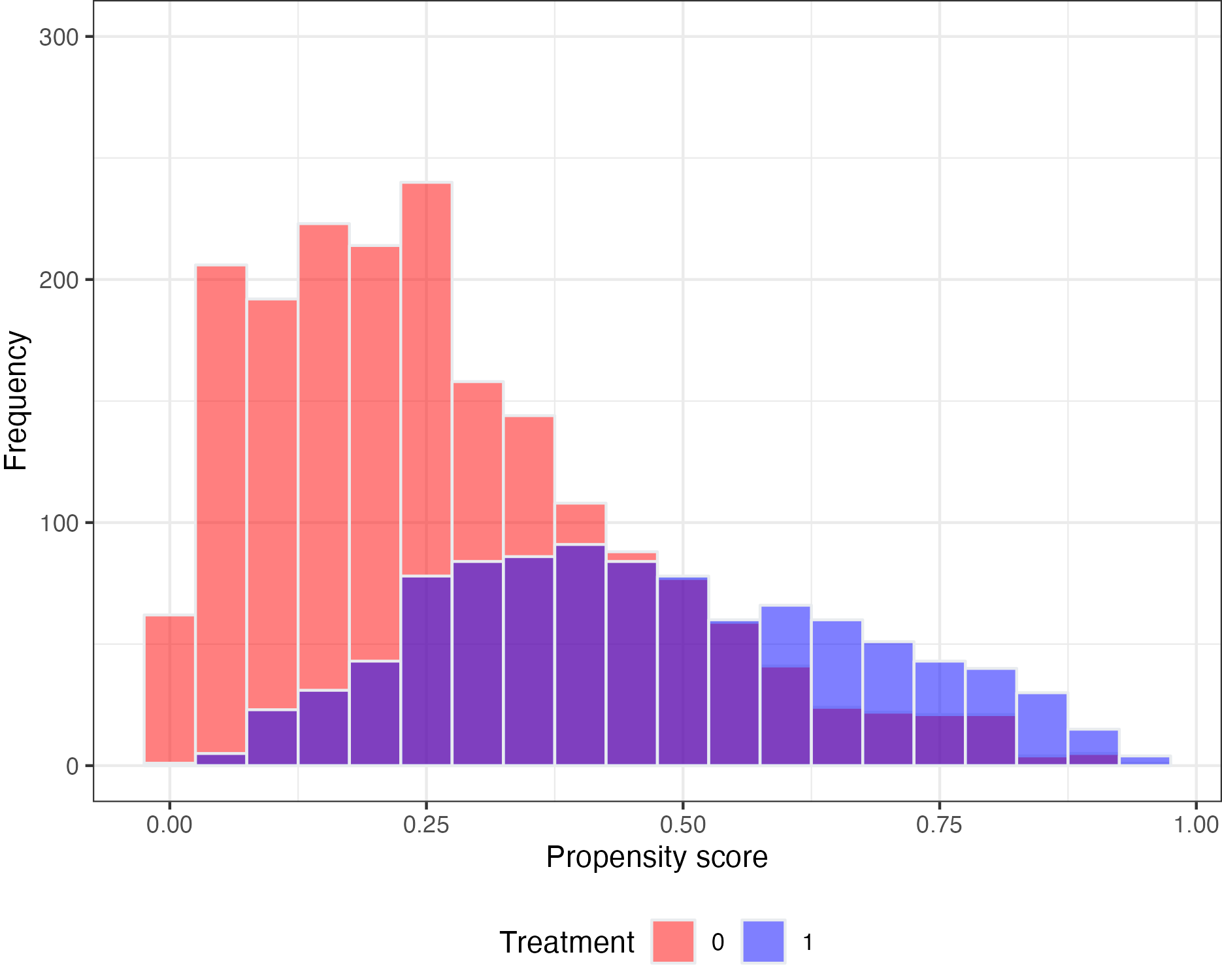}
\caption{Overlap of the case study data}
\label{fig:Overlap_real_data}
\end{figure}

\end{document}